%% file: ms.tex
\documentclass[iop]{emulateapj}
\usepackage[modulo]{lineno}
\usepackage[xcolor]{xcolor}
\usepackage{amsmath}
\usepackage[breaklinks,colorlinks,citecolor=blue]{hyperref} 
\usepackage{mathtools}

\def\deg{{$^{\circ}$}}
\def\bd{\object[BD-18 5550]{BD$-$18\deg5550}}
\def\csa{\object[BPS CS 22185-007]{CS~22185--007}}
\def\csb{\object[BPS CS 22891-200]{CS~22891--200}}

\def\msun{\mbox{$M_{\odot}$}}

\def\rpro{\mbox{$r$-process}}
\def\spro{\mbox{$s$-process}}

\def\ncap{\mbox{$n$-capture}}

\def\retgal{\object[NAME Reticulum II]{Ret~II}}

\def\booonegal{\object[NAME Bootes Dwarf Spheroidal Galaxy]{Boo~I}}
\def\bootwogal{\object[NAME Bootes II]{Boo~II}}
\def\cvntwogal{\object[NAME Canes Venatici II]{CVn~II}}
\def\comgal{\object[NAME Coma Dwarf Galaxy]{Com}}
\def\hergal{\object[NAME Her Dwarf Galaxy]{Her}}
\def\leogal{\object[NAME Leo IV Dwarf Galaxy]{Leo~IV}}
\def\segonegal{\object[NAME Segue 1]{Seg~1}}
\def\segtwogal{\object[NAME Segue 2]{Seg~2}}
\def\tuctwogal{\object[NAME Tucana II]{Tuc~II}}

\def\umagal{\object[NAME UMa II Galaxy]{UMa~II}}

\shorttitle{Heavy Metals in Ultra-Faint Dwarf Galaxies}
\shortauthors{I.U.\ Roederer}
\slugcomment{Accepted for publication in the Astrophysical Journal}

\begin{document}

\title{
The Origin of the Heaviest Metals in Most 
Ultra-Faint Dwarf Galaxies}

\author{Ian U.\ Roederer\altaffilmark{1,2}}

\altaffiltext{1}{Department of Astronomy, University of Michigan,
1085 S.\ University Ave., Ann Arbor, MI 48109, USA;
\mbox{iur@umich.edu}}

\altaffiltext{2}{Joint Institute for Nuclear Astrophysics and Center for the
Evolution of the Elements (JINA-CEE), USA}


\addtocounter{footnote}{2}

\begin{abstract}

The heaviest metals found in stars in
most ultra-faint dwarf (UFD) galaxies in the Milky Way halo
are generally underabundant by an order of magnitude or more
when compared with stars in the halo field.
Among the heavy elements produced by
$n$-capture reactions,
only Sr and Ba can be detected in 
red giant stars 
in most UFD galaxies.
This limited chemical information 
is unable to identify the 
nucleosynthesis process(es)
responsible for producing the heavy elements in
UFD galaxies.
Similar [Sr/Ba] and [Ba/Fe] ratios are found 
in three bright halo field stars,
BD$-$18$^{\circ}$5550, CS~22185--007, and CS~22891--200.
Previous studies of high-quality spectra of these stars
report detections of additional $n$-capture elements,
including Eu.
The [Eu/Ba] ratios in these stars span $+$0.41 
to $+$0.86.
These ratios and others among
elements in the rare earth domain
indicate an $r$-process origin.
These stars have some of the
lowest levels of $r$-process enhancement known,
with [Eu/H] spanning $-$3.95 to $-$3.32,
and they may be considered nearby 
proxies for faint stars in UFD galaxies.
Direct confirmation, however, must await
future observations of additional
heavy elements in stars in the UFD galaxies themselves.

\end{abstract}


\section{Introduction}
\label{intro}

Stars found in the lowest luminosity galaxies known
reveal the chemical composition of the molecular
clouds seeded by metals produced by
only one or a few prior generations of stars.
These ultra-faint dwarf (UFD) galaxies 
date from the early Universe
(e.g., \citealt{brown14}).
The elements found within their stars
inform our understanding of the
first stars, the first metal production,
the process of galaxy formation,
and the nature of the building blocks of our own Milky Way.

High-resolution optical spectra have been obtained
and analyzed for individual stars in 11~UFD
galaxies around the Milky Way.  
These galaxies range from 23 to 160~kpc from the Sun
\citep{mcconnachie12,bechtol15},
and the faintness of their brightest red giant stars
(typically 16~$< V <$~19)
limits the quantity and quality of high-resolution
spectra that can be obtained.
Even so,
elements heavier than the iron group---Sr ($Z =$~38)
or Ba ($Z =$~56)---have been detected 
in all but one 
(\bootwogal; \citealt{francois16,ji16a})
of these galaxies
and nearly all halo field stars that have been studied
\citep{roederer13}.
As detailed chemical abundances have been
presented for stars in more UFD galaxies, a
consistent, if yet unexplained, pattern has emerged.
In general,
the heaviest elements are underabundant
relative to Fe or H when compared with
halo field stars at similar metallicities
(e.g., \citealt{koch13,frebel15}).

Other heavy elements are rarely detected
in stars in the UFD galaxies.
These elements could be present yet remain undetectable
because
their abundances are even lower than Sr and Ba
and their electronic transitions are not concentrated
in a few strong lines in the same way that
Sr~\textsc{ii} and Ba~\textsc{ii} are
(e.g., \citealt{roederer13}).
\citet{frebel10} made the first detection of 
La ($Z =$~57) in any UFD,
and 
\citet{frebel14} made the first detections of 
elements heavier than La in any UFD.
These detections were 
possible because the stars in question
probably were enriched after their birth
by \spro\ material 
from a more evolved
binary companion star that passed through the
asymptotic giant branch (AGB)
phase of evolution.

It was not until the detection of high levels of
\rpro\ enhancement in many stars in \retgal\
that elements heavier than Ba were detected
in any UFD star whose present-day composition
reflects its natal composition 
\citep{ji16b,ji16c,roederer16a}.
In the case of \retgal,
the nucleosynthesis process responsible for the
production of the heavy elements in these stars
is easily identified as the \rpro\ 
because
so many elements are detectable.
\citet{ji16b} conclude that any rare site that produces
a large yield of \rpro\ material is 
consistent with the constraints imposed by \retgal.~
\citet{lee13} anticipated this outcome, 
and their chemical evolution models
assumed a strong mass-dependent yield of \rpro\ material
from a small fraction of core-collapse supernovae.
Even if the astrophysical site associated with the
\rpro\ material in \retgal\ cannot be unambiguously identified,
identifying an environment teeming with \rpro\ material
is an important step toward this goal.

Setting \retgal\ aside, 
the stars in UFD galaxies with low levels of \ncap\ elements
are distinct 
in their [Sr/Ba] and [Ba/Fe] ratios
relative to the majority of halo field stars
\citep{ji16c}.
\citet{frebel15} surmised that 
these low, distinct heavy-element abundances
could be a signature of the earliest star-forming clouds.
If so, this raises the tantalizing possibility
of identifying the first \ncap\ process
to have operated in the early Universe,
perhaps even in the first stars.

What was that process?
At present, the answer is unclear.
Other work on candidate 
second-generation stars in the halo field
implicates some form of \rpro\ nucleosynthesis
\citep{roederer14a}.
Only one ratio---[Sr/Ba]---is available
in the stars in UFD galaxies, so 
it is difficult to exclude candidate processes.
In principle, it should be possible to
detect additional heavy elements in the spectra
of individual metal-poor red giant stars in the UFD galaxies
with higher signal-to-noise (S/N) spectra.
In practice, obtaining such spectra is extremely challenging 
because
large-aperture telescope time is limited and
these stars are so faint.
A more tractable approach could be to 
examine nearby, bright halo field stars.
Additional \ncap\ elements may be
detectable in the high S/N spectra of the few
bright halo field stars that occupy the same
region of chemical space as the stars in most UFD galaxies.
Here, I identify three such stars with low levels
of \ncap\ elements.
The chemical composition of these halo field stars
is likely to be similar to the stars in most UFD galaxies.
These stars may represent our best opportunity to identify
the nucleosynthesis process(es) responsible for producing
the \ncap\ elements found in extremely low levels
in the UFD galaxies.

\section{Literature Sample}
\label{sample}

\begin{figure*}
\begin{center}
\includegraphics[angle=0,width=5.0in]{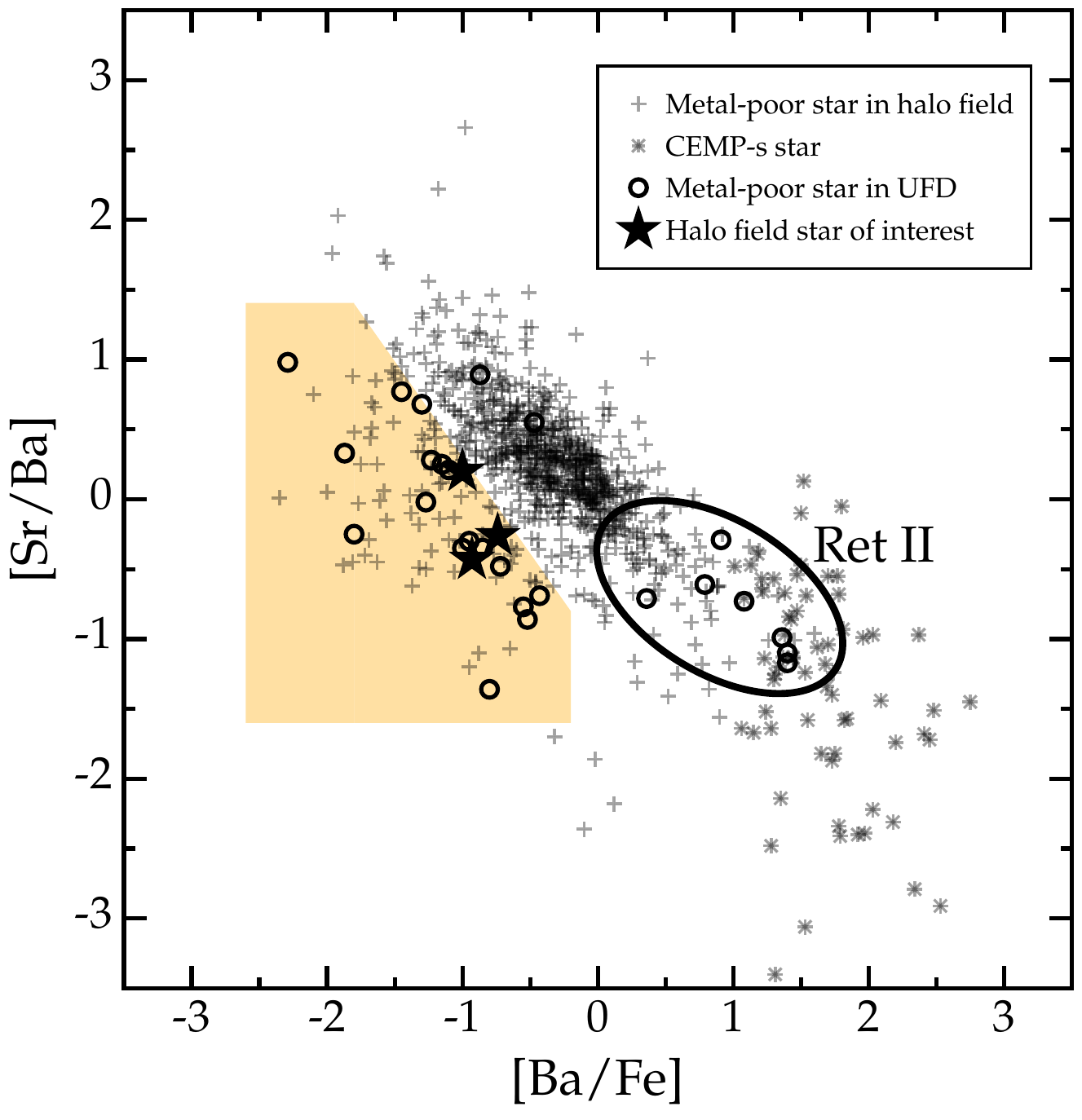}
\end{center}
\caption{
\label{srbaplot}
Sr and Ba abundances in
field stars and stars in UFD galaxies.
The crosses mark individual 
metal-poor ([Fe/H]~$< -$~1.5) stars in the halo field,
the circles mark individual metal-poor stars in UFD galaxies,
and the large star-shaped symbols mark
the three field stars of interest:\
\mbox{BD$-$18$^{\circ}$5550},
\mbox{CS~22185--007}, and \mbox{CS~22891--200}.
Halo stars marked with an eight-pointed shape are suspected to 
have received large amounts of carbon and \spro\ material
(CEMP-$s$ class; \citealt{beers05})
from a companion star that passed through the 
AGB phase of evolution.
The large oval identifies the location of
the seven known highly \rpro-enhanced stars
in the \mbox{Ret~II} UFD.~
The shaded box marks the location
where this study seeks to identify
field stars in the same region of the diagram
occupied by the stars in UFD galaxies
with low levels of \ncap\ elements.
Upper limits have been omitted for clarity.
}
\end{figure*}

Figure~\ref{srbaplot} 
illustrates the [Sr/Ba] and [Ba/Fe] ratios
for 977~metal-poor 
([Fe/H]~$< -$1.5) stars
in the halo field 
(951~stars) and UFD galaxies
(26~stars).
The literature sources are listed in Table~\ref{littab}.
Duplicates have been removed from the sample.
The abundance data
for the field stars are taken mainly
from recent large surveys, supplemented
with data from studies of individual stars.
Consequently, this 
is not an unbiased sample.
Nevertheless, it is 
useful for the purpose of the present study
simply because it populates the
region of interest in Figure~\ref{srbaplot}.

The shaded region in Figure~\ref{srbaplot} 
highlights the parameter space where
most stars in UFD galaxies are found.
This region spans several dex in [Sr/Ba],
and it is offset to lower [Ba/Fe] by 
more than 1~dex from the main locus of
halo field stars.
Each of the 
85~halo field stars found in this region
is a candidate for further consideration.
I require that both Sr and Ba be detected
in a given star.
This is a practical choice, 
because
stars with no detectable Sr or Ba are unlikely
to have detections from weaker lines
of less abundant elements.

\input{tab1}

Most of the stars (seven of nine) 
that have been studied in the \retgal\ UFD galaxy
are highly enhanced in \rpro\ material,
unlike the remaining two stars in \retgal\ 
and stars in the other 10~UFD galaxies that have been studied.
The seven \rpro-enhanced stars in \retgal\
are 
circled with the bold line
in Figure~\ref{srbaplot}.

Detailed abundances have been studied in
three other UFD galaxies that are not illustrated in Figure~\ref{srbaplot},
\bootwogal, \cvntwogal, and \segonegal.
Sr and Ba have been detected in \segonegal,
but not simultaneously in any given star
with [Fe/H]~$< -$1.5
\citep{frebel14}.
One star has been studied in \cvntwogal,
and only Sr has been detected there \citep{francois16}.
At present, only \bootwogal\ lacks any 
compelling detection of Sr or Ba,
though only four stars in \bootwogal\ have been examined
\citep{koch14,francois16,ji16a}.
Upper limits on Sr and Ba in stars in 
\bootwogal, \cvntwogal, and \segonegal\
suggest they occupy similar regions of parameter space
as the stars in other UFD galaxies
(e.g., \citealt{frebel14,ji16c}).

\section{Results}
\label{results}

\begin{figure*}
\begin{center}
\includegraphics[angle=0,width=3.4in]{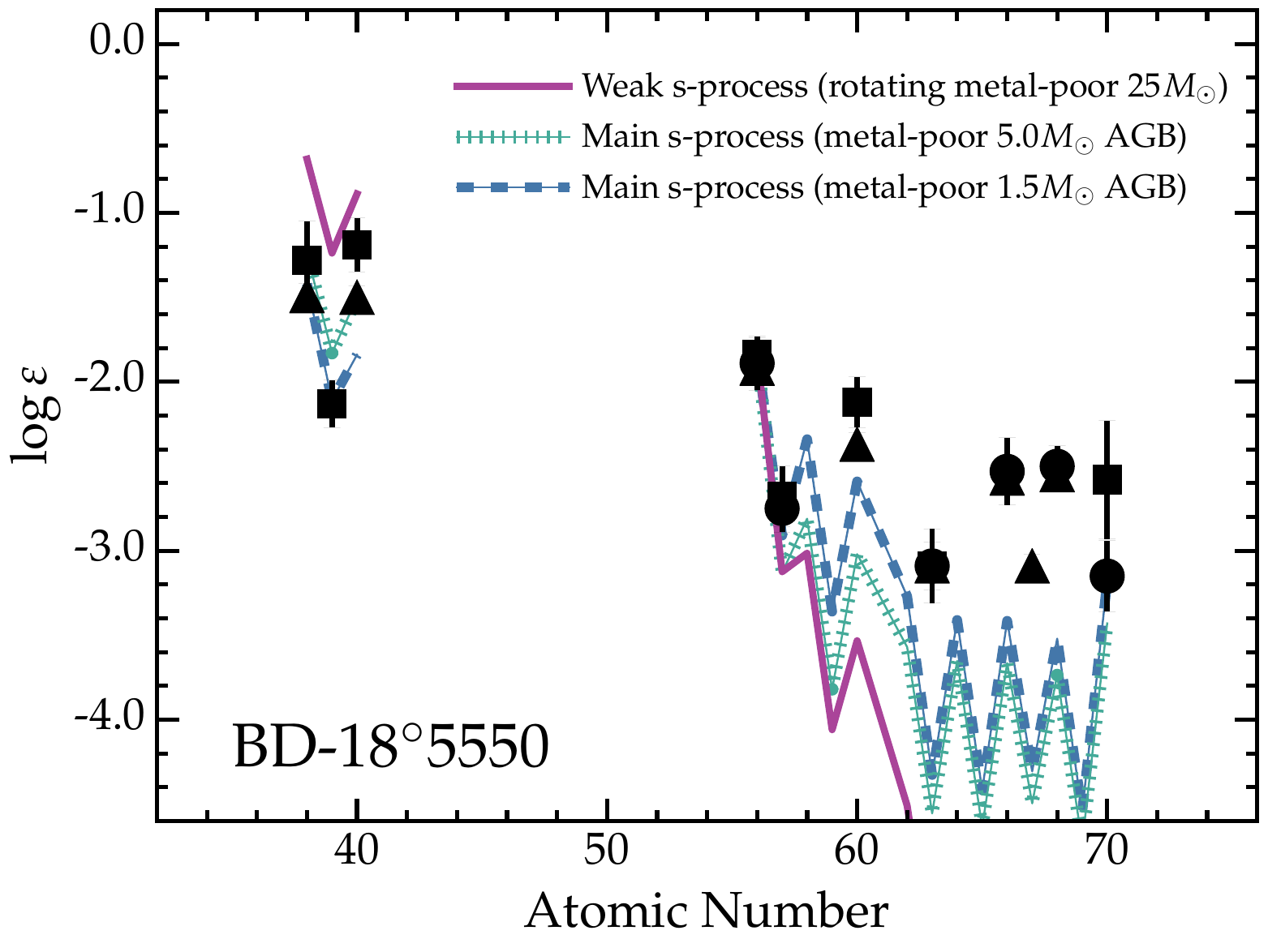} 
\hspace*{0.1in}
\includegraphics[angle=0,width=3.4in]{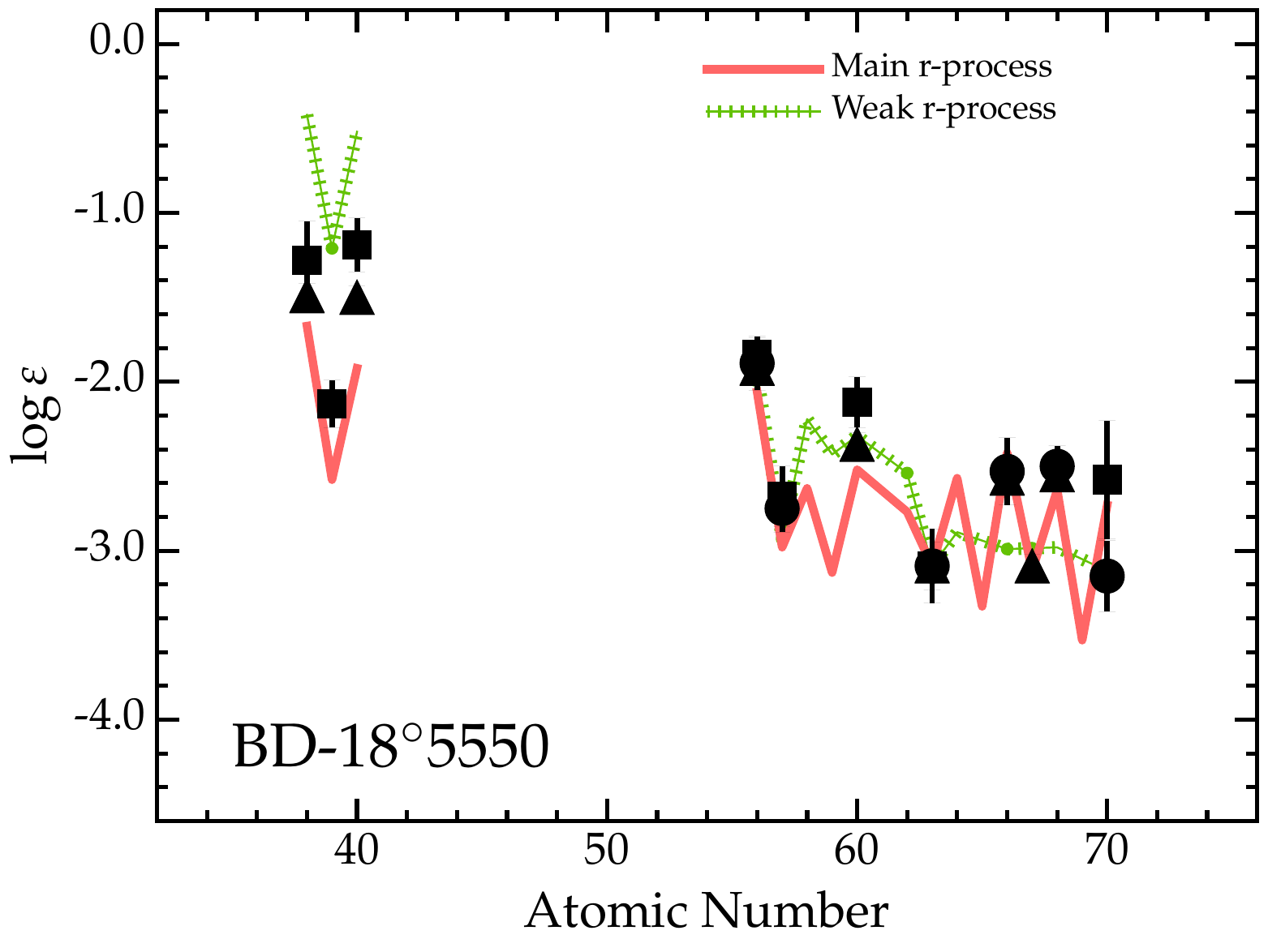} \\
\includegraphics[angle=0,width=3.4in]{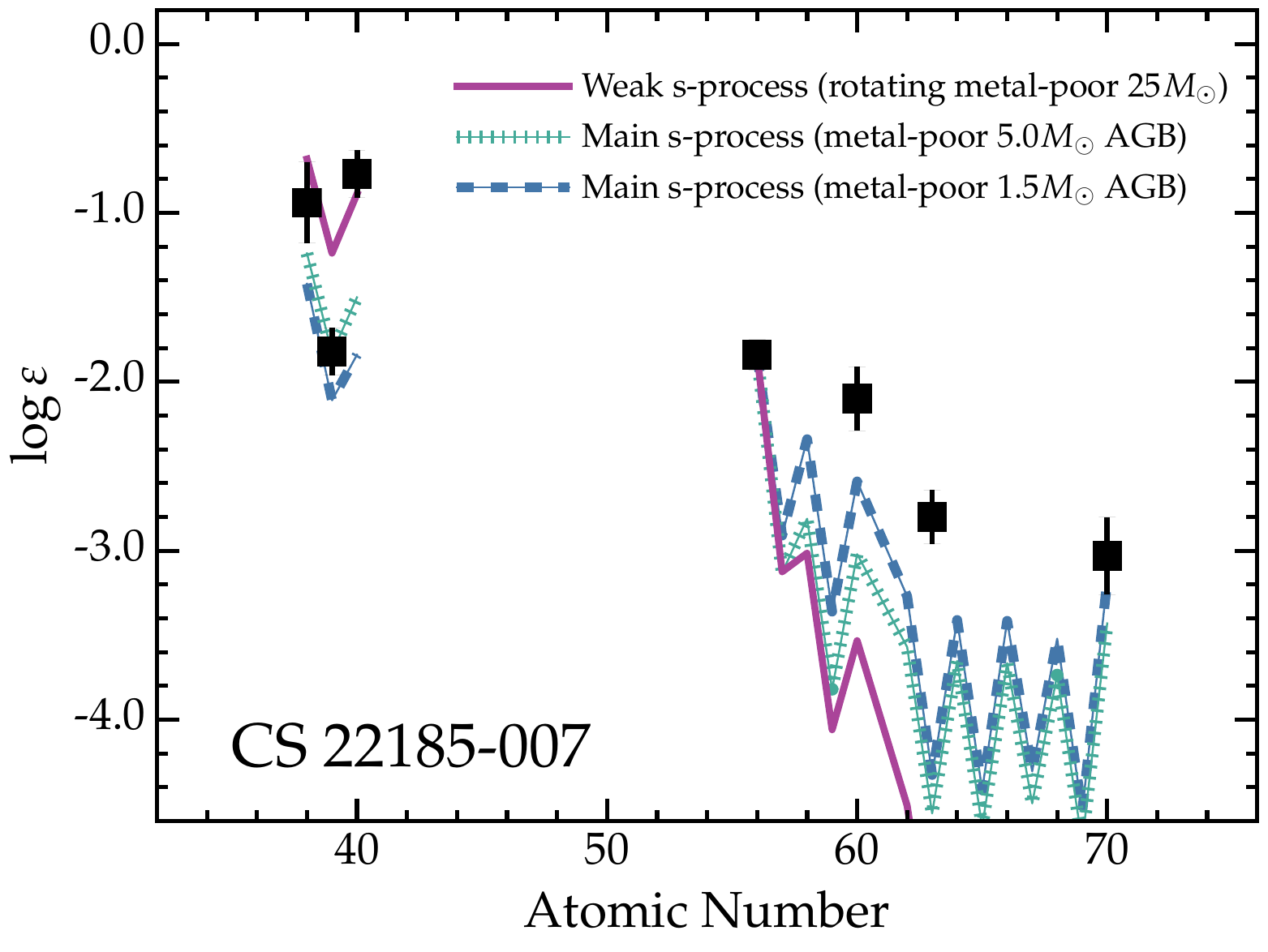}
\hspace*{0.1in}
\includegraphics[angle=0,width=3.4in]{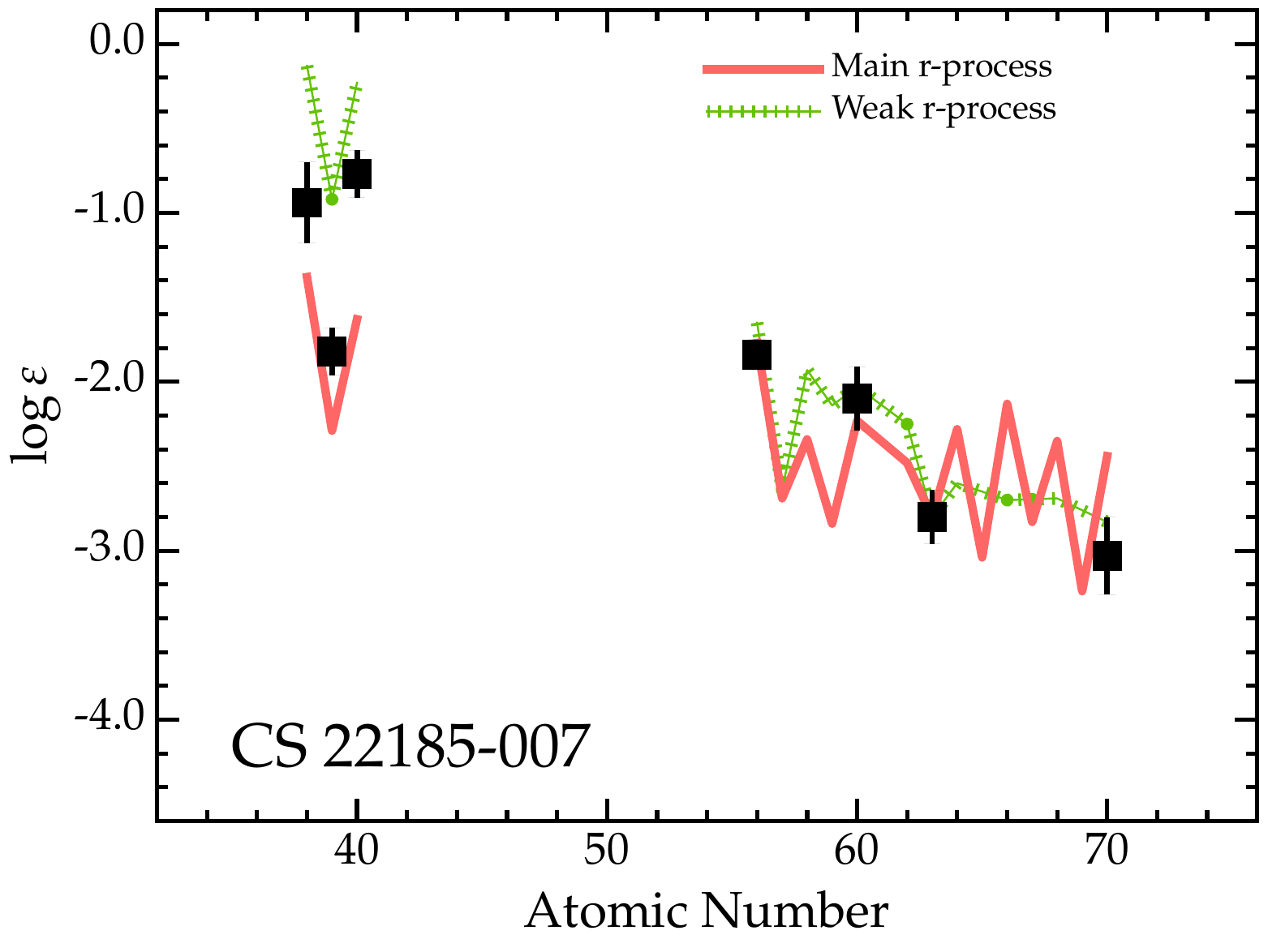} \\
\includegraphics[angle=0,width=3.4in]{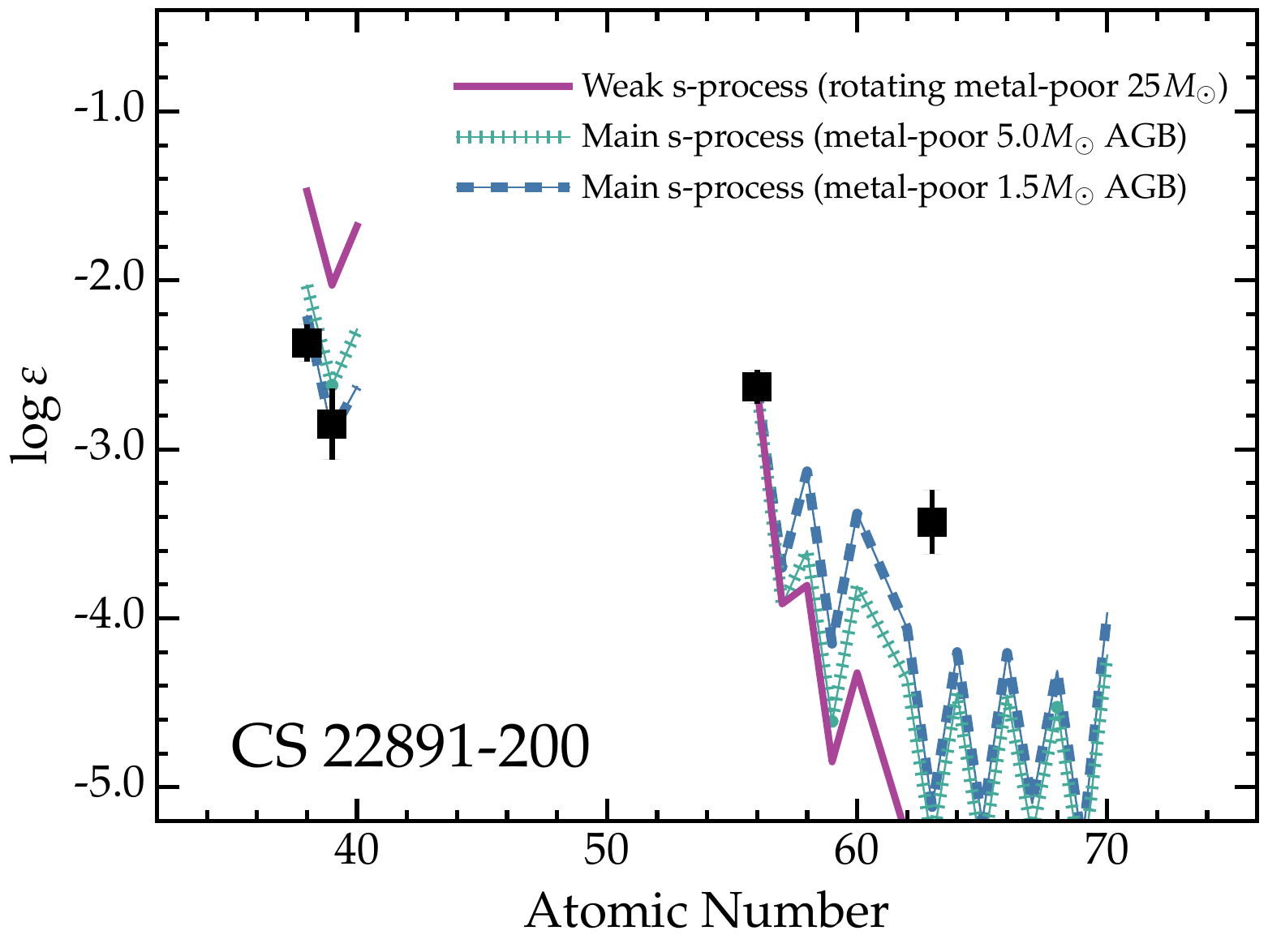} 
\hspace*{0.1in}
\includegraphics[angle=0,width=3.4in]{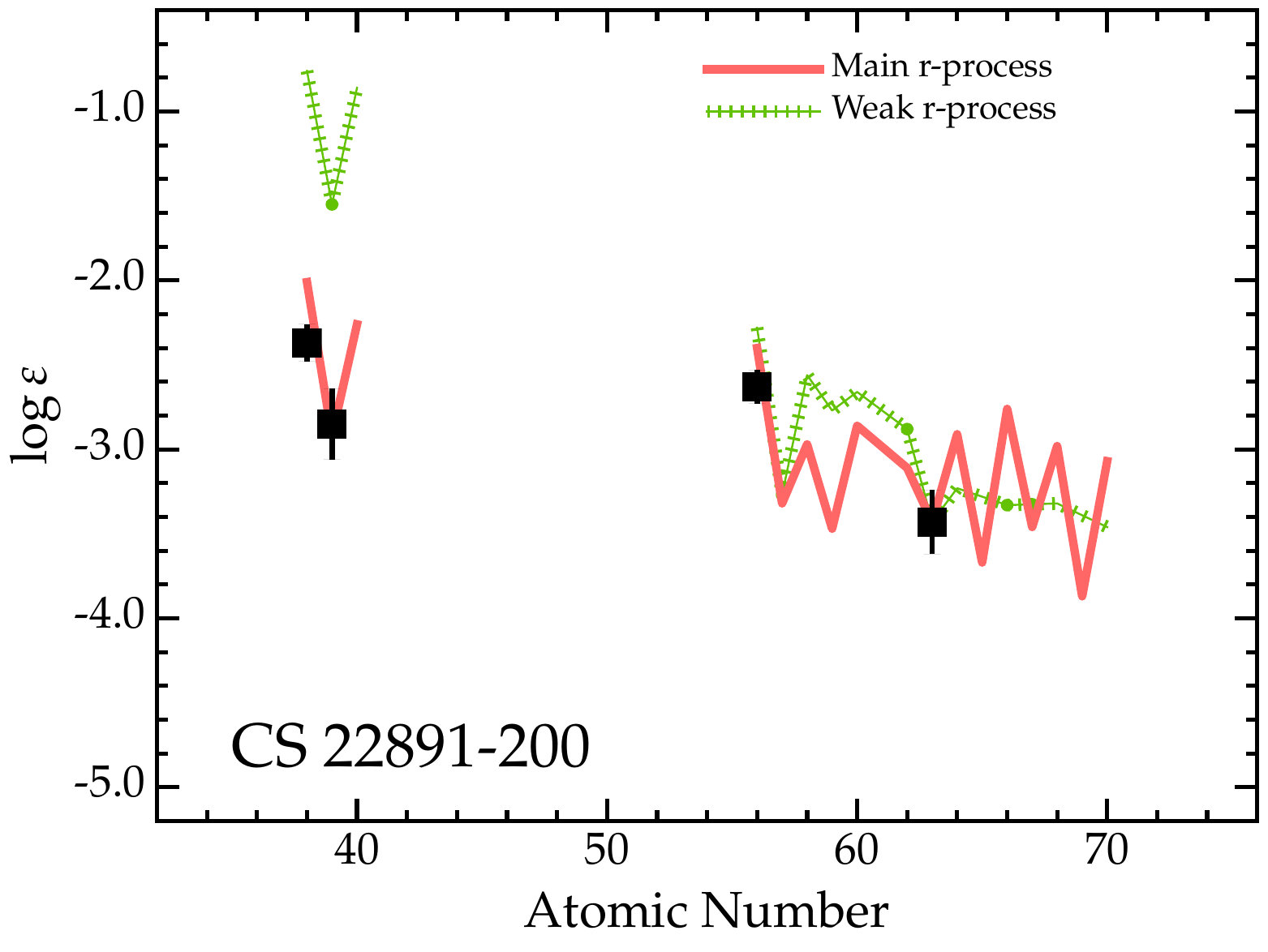} \\
\end{center}
\caption{
\label{logepsplot}
Log of the abundance as a function of atomic number.
Three abundance studies are illustrated:\
the circles mark data from \citet{johnson01},
the triangles mark data from \citet{francois07},
and
the squares mark data from \citet{roederer14c}.
The abundances for the three different studies of
\mbox{BD$-$18$^{\circ}$5550} have been normalized
to the Eu abundance from \citeauthor{roederer14c} 
In the left panels,
the solid purple line indicates the template for the 
weak component of the \spro\ operating in 
massive, low-metallicity,
rapidly-rotating stars \citep{frischknecht12,frischknecht16},
the short-studded aqua line indicates the template for the
main component of the \spro\ operating in a 
metal-poor 5.0~\msun\ AGB star,
and the long-studded blue line indicates the template for the
main component of the \spro\ operating in a
metal-poor 1.5~\msun\ AGB star with the same parameters
\citep{cristallo11,cristallo15}.
These are normalized to the Ba abundance.
In the right panels,
the solid red line indicates the template for the
main component of the \rpro\ and
the short-studded green line indicates the template for the
weak component of the \rpro.
These are normalized to the Eu abundance.
See Section~\ref{results} for details.
}
\end{figure*}

Most halo field stars 
in the shaded region of Figure~\ref{srbaplot}
lack any additional detections of heavy elements
other than Sr and Ba, as might be expected 
given the low abundances.
A few stars show detections of Y ($Z =$~39) or Zr ($Z =$~40),
but these elements offer little new information
that could be used to 
distinguish the nucleosynthesis mechanism
responsible for producing the heavy elements.
Only three stars show compelling detections
of one or more elements heavier than Ba:\
\bd, \csa, and \csb.
Their heavy element abundance patterns 
are illustrated in Figure~\ref{logepsplot}.
The fact that these three stars lie near the right side of
the shaded region is not surprising,
because
detection becomes
increasingly difficult as the overall abundances
decrease from right to left.

\bd\ is a bright ($V \approx$~9.3) 
metal-poor ([Fe/H]~$\approx -$3.0)
red giant that has been the subject
of many studies since its discovery by \citet{bond80}.
Three studies in the current century 
\citep{johnson01,francois07,roederer14c}
have reported detections of several \ncap\ elements
in addition to Sr and Ba.
These abundances are shown in Figure~\ref{logepsplot},
where they have been normalized to the Eu abundance
to account for small systematic shifts
in the overall abundance scale.
The $\log$(X/Eu) ratios between a given element, X, and Eu
agree among different studies within $\sim 2\sigma$.
The level of \ncap\ abundances found in \bd\
is among the lowest known:\
$\log\epsilon$(Eu)~$= -$3.09~$\pm$~0.18;
[Eu/H]~$= -$3.61~$\pm$~0.18;
[Eu/Fe]~$= -$0.46~$\pm$~0.14 \citep{roederer14c}.

\csa\ is somewhat fainter ($V \approx$~13.4) than \bd,
but it is still considerably brighter than the brightest
red giants in UFD galaxies.
This metal-poor ([Fe/H]~$\approx -$3.0) giant was first 
identified by \citet{beers92},
and \citet{roederer14c} performed the first
abundance analysis on a high-resolution spectrum of this star.
That study reported detections of 
one line of Nd~\textsc{ii}, 
two lines of Eu~\textsc{ii}, and 
one line of Yb~\textsc{ii}.
The abundances of \ncap\ elements in \csa\ are also quite low:\
$\log\epsilon$(Eu)~$= -$2.80~$\pm$~0.20;
[Eu/H]~$= -$3.32~$\pm$~0.20;
[Eu/Fe]~$= -$0.31~$\pm$~0.16 \citep{roederer14c}.

\csb\ is similarly faint ($V \approx$~13.9).
This metal poor ([Fe/H]~$\approx -$3.9) 
giant was first identified by \citet{beers85}.
\citet{mcwilliam95} performed the first detailed abundance analysis,
and subsequent abundance work has been performed by 
\citet{mcwilliam98}, \citet{andrievsky11},
\citet{hollek11}, and \citet{roederer14c}.
The highest S/N spectrum of \csb\ was examined by \citeauthor{roederer14c},
who reported a detection of Eu from
the two strongest Eu~\textsc{ii} lines in the blue.
Eu is present in \csb\
at an extremely low level:\
$\log\epsilon$(Eu)~$= -$3.43~$\pm$~0.22;
[Eu/H]~$= -$3.95~$\pm$~0.22;
[Eu/Fe]~$= -$0.07~$\pm$~0.19 \citep{roederer14c}.

Three lines are shown for comparison in the left panels 
of Figure~\ref{logepsplot}.
One is a theoretical prediction for the
weak component of the \spro\ operating in a 25~\msun,
rapidly-rotating (at half the critical
break-up velocity), low-metallicity 
($Z = 10^{-5}$, [Fe/H]~$\approx -$3.2) star
(G.\ Cescutti 2016, private communication\footnote{
\url{http://www.astro.keele.ac.uk/shyne/datasets/s-process-yields-from-frischknecht-et-al-12-15}};
\citealt{frischknecht12,frischknecht16}).
The other two are theoretical predictions for 
the main component of the \spro\
operating in 5.0 and 1.5~\msun\ 
AGB stars with $Z = 10^{-4}$ ([Fe/H]~$\approx -$2.2,
the lowest metallicity available in the grid).
These two are taken from the Full-Network Repository 
of Updated Isotopic Tables and Yields (FRUITY)
database\footnote{
\url{http://fruity.oa-teramo.inaf.it/modelli.pl}}
\citep{cristallo11,cristallo15}.
It is unlikely that star formation 
lasted long enough in UFD galaxies to incorporate
yields from low- or intermediate-mass AGB stars, but
these three predictions represent the range of
\spro\ nucleosynthesis outcomes that could be expected
from low-metallicity stars.
It is immediately apparent from Figure~\ref{logepsplot}
that all of the \spro\ templates are a poor representation 
of the Eu and heavier elements in the rare earth domain
in these three stars.
Changing the normalization point for the models
does not affect this conclusion.

Two lines are shown for comparison in the right panels 
of Figure~\ref{logepsplot}.
One is a template for the main component of the \rpro,
which mirrors the solar system \rpro\ residual pattern.
This is formed by averaging the abundance patterns
in the highly \rpro-enhanced red giants
\object[BPS CS 22892-052]{CS~22892--052}
and 
\object[BPS CS 31082-001]{CS~31082--001}
\citep{hill02,sneden03a,sneden09}.
Another is a template for the weak component of the \rpro,
which may result from an inefficient or incomplete \rpro\
whose neutron flux is insufficient to flow to the
heaviest \ncap\ elements
(e.g., \citealt{truran02}).
This template is
formed by averaging together the abundance patterns
in the red giants
\object[HD 88609]{HD~88609} 
and
\object[HD 122563]{HD~122563}
\citep{honda07}.
The templates for the main and weak
components of the \rpro\ are derived from other metal-poor 
red giant stars,
so any effects due to, e.g., 
departures from local thermodynamic equilibrium should cancel
when performing this relative comparison.

The two \rpro\ templates provide much better
representations of the observed abundance patterns,
especially throughout the rare earth domain 
(Ba, La, Nd, Eu, Dy, Ho, Er, and Yb) in \bd\ and \csa.
There may be a preference in the \bd\ data for the 
main component of the \rpro;
e.g., at Dy, Ho, and Er ($Z =$~66--68).
The \csa\ data may favor the weak component of the \rpro,
especially at Yb ($Z =$~70).
The Sr, Y, and Zr ($Z =$~38--40) 
abundances in \bd\ and \csa\
lie between the main and weak \rpro\ patterns
when normalized to Eu.
The abundance pattern in \csb, though limited,
favors the main component of the \rpro.
The elements in the rare earth domain
in these stars are more than 2~orders of magnitude 
less abundant than in highly \rpro-enhanced stars like 
\object[BPS CS 22892-052]{CS~22892--052}
or those in \retgal,
yet they are still recognizable as having
originated via some form of \rpro\ nucleosynthesis.

The [Eu/Ba] ratio is a good quantitative discriminant of 
the nucleosynthetic origin of the elements in the
rare earth domain.
The [Eu/Ba] ratios in these three stars are supersolar:\
$+$0.41~$\pm$~0.16,
$+$0.69~$\pm$~0.17, and
$+$0.86~$\pm$~0.20 in
\bd, \csa, and \csb, respectively.
For comparison, \citet{roederer14d} found an average
[Eu/Ba]~$= +$0.71 with a star-to-star dispersion of
0.19~dex for 13 highly \rpro-enhanced stars drawn from the
same survey and analyzed in an identical fashion.
This is comparable to the [Eu/Ba] ratio expected from
the solar \rpro\ residuals, $+$0.92
(e.g., \citealt{sneden08,bisterzo11}).
The two stars frequently associated with the
weak component of the \rpro, 
\object[HD 88609]{HD~88609} and 
\object[HD 122563]{HD~122563}, 
have [Eu/Ba]~$= +$0.48~$\pm$~0.16 and
$+$0.53~$\pm$~0.18 \citep{honda07}
or [Eu/Ba]~$= +$0.33~$\pm$~0.16
and $+$0.33~$\pm$~0.18 \citep{roederer14c}.
The solar (main) \rpro\ and weak \rpro\ values
bracket the range of values found in the three stars of interest.
In contrast,
the [Eu/Ba] ratio expected from the \spro\ contribution
to the solar system is $-$1.17.
The ratios in these stars clearly prefer the \rpro\ values.

None of the light element (C to Zn; 6~$\leq Z \leq$~30) 
abundances in \bd, \csa, or \csb\
are unusual with respect to the metal-poor halo field population
\citep{roederer14a,roederer14c}.
\bd\ and \csa\ show enhancements in $\alpha$ elements
like Mg, Si, and Ca 
that are typical for metal-poor stars
([$\alpha$/Fe]~$\approx +$0.4;
e.g., \citealt{mcwilliam95,cayrel04}).
\csb\ is enhanced in 
C ([C/Fe]~$\approx +$1.0 after 
correcting for evolutionary effects; \citealt{placco14}),
N ([N/Fe]~$= +$1.2), 
and several $\alpha$ elements
(e.g., [Mg/Fe]~$= +$0.8).
These enhancements are common among stars in the 
class of carbon-enhanced metal-poor stars
with no enhancement of neutron-capture elements
(e.g., \citealt{ryan05,norris13}).
Several stars have been found that
show both carbon enhancement and \rpro\ enhancement
(e.g., \citealt{sneden03a,roederer14c,ji16c}).
The iron-group elements in these stars show no deviations
from the usual ratios, either.
In summary, the light-element abundance patterns
in these stars are typical
among metal-poor stars in the halo and UFD galaxies.

\section{Discussion and Conclusions}
\label{discussion}

Among 85~metal-poor halo field stars with
detections of Sr and Ba
in the shaded region of Figure~\ref{srbaplot},
only 3 of them have compelling detections of
additional elements heavier than Ba.
There is no known 
association between these field stars
and disrupted satellite galaxies.
These stars could have formed in UFD
galaxies that were later disrupted, 
but the result of this study 
does not rely on such speculation.
These stars
provide the first definitive evidence
than \rpro\ nucleosynthesis can produce
ratios consistent with the [Sr/Ba] and [Ba/Fe]
ratios found in most UFD galaxies.

As can be seen in Figure~\ref{srbaplot}, and 
which was found previously by 
\citet{frebel16} and \citet{ji16c}, 
the [Sr/Ba] ratios in most UFD galaxies
span a range of several dex.
This range overlaps with the [Sr/Ba] ratios 
found in the highly \rpro-enhanced stars in \retgal,
hinting that the heavy elements in the other UFD galaxies
could reflect the same \rpro\ pattern diluted into
a larger mass of Fe.
Two of the three field stars examined here,
\bd\ and \csb,
also fall into this range.
A substantial fraction of stars in the UFD galaxies
and the field star \csa\ have solar or super-solar [Sr/Ba] ratios
that are inconsistent with
the \rpro\ enhanced stars in \retgal.
This indicates that an additional
nucleosynthesis mechanism may be responsible.
This additional process could correspond to the weak
\rpro\ (e.g., \citealt{wanajo13}),
truncated \rpro\ \citep{boyd12,aoki13b}, or
weak \spro\ (e.g., \citealt{frischknecht16}).
If a weak or truncated \rpro\ is responsible,
it would need to be capable of 
producing variable [Sr/Ba] yields
and relatively normal \rpro\ ratios within the rare earth domain.
If a weak \spro\ is responsible,
it would need to be capable of producing
variable [Sr/Ba] yields,
and its products would need to be mixed
with a small amount of \rpro\ material
before being incorporated into the stars observed today.

\citet{cescutti13} were able to 
reproduce the spread in [Sr/Ba] ratios
in metal-poor halo stars 
using a combination of 
a main \rpro\ component from supernovae and
a weak \spro\ component from 
massive, rapidly-rotating, low-metallicity stars.
Their model predicts that the stars with
high [Sr/Ba] ratios should show 
\spro\ signatures.
\citet{cescutti14} and 
\citet{cescutti15} 
considered other sites for the \rpro\ nucleosynthesis,
including
the mergers of binary neutron star systems
and magneto-rotational supernovae.
These models predict that the stars with low [Sr/Ba]
ratios should show \rpro\ signatures,
in agreement with observations.
These models also predict that metal-poor stars with high
[Sr/Ba] ratios should show \spro\ signatures,
independent of the site of the \rpro.
The results of the present study 
suggest an observational test of these models:\
derivation of the abundance pattern within the rare earth
domain in stars with [Sr/Ba]~$>$~0
and [Ba/Fe]~$< -$1.0 or so.
The \citet{cescutti15} models 
may be capturing several important pieces of physics, like
the mass range of stars where the \ncap\ nucleosynthesis operates
or the 
typical mass of H or Fe into which the \ncap\ elements are diluted,
for example.
Future theoretical work should address whether these scenarios
can also be applied to environments that
give rise to today's population of UFD galaxies
(cf.\ \citealt{tsujimoto14,ishimaru15}).
Future theoretical work should also address whether 
the massive, rapidly-rotating, low-metallicity stars that are proposed to
produce a weak \spro\ might also produce 
a small amount of \rpro\ material during the
subsequent supernova explosion.

Certainly the most direct observational approach
to test these assertions
is to detect additional \ncap\ elements
in stars in the UFD galaxies themselves.
This may require the use of high-resolution 
echelle spectrographs
on 20--30~m class telescopes,
like the G-CLEF instrument \citep{szentgyorgyi14}
being developed for the
Giant Magellan Telescope.
In the meantime,
the results presented here
may offer the next-best 
observational guidance
for interpreting the heavy element abundance
patterns in most UFD galaxies.

\acknowledgments

I thank G.\ Cescutti, J.\ Cowan, and A.\ Ji
for supportive comments on an early version 
of this manuscript, and 
I thank G.\ Cescutti for sending 
model predictions in tabular form.
I also thank the referees,
G.\ Preston and an anonymous referee,
for their helpful recommendations.
I acknowledge partial support from
grant PHY~14-30152 (Physics Frontier Center/JINA-CEE)
awarded by the U.S.\ National Science Foundation. 
This research has made use of NASA's
Astrophysics Data System Bibliographic Services;
the arXiv pre-print server operated by Cornell University;
the SIMBAD and VizieR
database hosted by the
Strasbourg Astronomical Data Center;
and
the \texttt{matplotlib} \citep{hunter07}, 
\texttt{numpy} \citep{vanderwalt11}, and 
\texttt{scipy} \citep{jones01}
Python libraries.

\end{document}

%% file: tab1.tex
\begin{deluxetable}{ccc}
\tablecaption{Literature Sources for Abundance Data
\label{littab}}
\tablewidth{0pt}
\tabletypesize{\scriptsize}
\tablehead{
\colhead{Reference} &
\colhead{No.\ stars} &
\colhead{Population} }
\startdata                                                                      
\citet{aoki02}            &  2 & field \\
\citet{aoki04}            &  1 & field \\
\citet{aoki05}            & 14 & field \\
\citet{aoki06}            &  1 & field \\
\citet{aoki10}            &  2 & field \\
\citet{aoki13a}           & 35 & field \\
\citet{barklem05}         &200 & field \\
\citet{bonifacio09}       &  5 & field \\
\citet{burris00}          &  9 & field \\
\citet{carretta02}        &  5 & field \\
\citet{cohen08}           &  1 & field \\
\citet{cohen13}           & 80 & field \\
\citet{depagne00}         &  1 & field \\
\citet{francois07}        & 30 & field \\
\citet{francois16}        &  2 & \hergal \\
\citet{frebel10}          &  6 & \comgal, \umagal \\
\citet{frebel16}          &  2 & \booonegal \\
\citet{gratton94}         &  2 & field \\
\citet{hansen11}          &  1 & field \\
\citet{hansen15}          & 15 & field \\
\citet{hollek11}          & 16 & field \\
\citet{honda04}           &  8 & field \\
\citet{honda11}           &  1 & field \\
\citet{ishigaki14}        &  2 & \booonegal \\
\citet{ivans03}           &  1 & field \\
\citet{jacobson15}        &115 & field \\
\citet{ji16c}             &  7 & \retgal \\
\citet{ji16d}             &  4 & \tuctwogal \\
\citet{johnson02}         & 13 & field \\
\citet{lai07}             & 53 & field \\
\citet{lai08}             & 21 & field \\
\citet{mashonkina10}      &  1 & field \\
\citet{mashonkina14}      &  1 & field \\
\citet{mcwilliam98}       &  3 & field \\
\citet{mishenina01}       &  6 & field \\
\citet{norris10}          &  1 & \booonegal \\
\citet{placco15}          &  1 & field \\
\citet{placco16}          &  1 & field \\
\citet{preston00}         & 14 & field \\
\citet{preston01}         &  2 & field \\
\citet{preston06}         &  6 & field \\
\citet{roederer10}        &  1 & field \\
\citet{roederer14b}       &  1 & \segtwogal \\
\citet{roederer14c}       &259 & field \\
\citet{roederer16b}       &  1 & field \\
\citet{roederer12}        &  1 & field \\
\citet{ryan91}            &  3 & field \\
\citet{ryan96}            &  2 & field \\
\citet{simon10}           &  1 & \leogal \\
\citet{sivarani06}        &  2 & field \\
\citet{sneden03b}         &  2 & field \\
\citet{yong13}            & 13 & field
\enddata
\end{deluxetable}